\documentclass[aps,prd,twocolumn,10pt,superscriptaddress,nofootinbib,nobibnotes,longbibliography]{revtex4-1}

\usepackage{amssymb}
\usepackage{graphicx}
\usepackage{amsmath}
\usepackage{hyperref}
\usepackage{subfigure}
\usepackage{multirow}
\usepackage{setspace}
\usepackage{verbatim}
\usepackage{float}
\usepackage{color}
\usepackage{ulem}
\usepackage[utf8]{inputenc}
\usepackage[table,xcdraw]{xcolor}
\usepackage{makecell}
\usepackage{url}

\begin{document}

\title{The DESI DR1/DR2 evidence for dynamical dark energy is biased by low-redshift supernovae}

\author{Lu Huang}
\email{huanglu@itp.ac.cn}
\affiliation{Institute of Theoretical Physics, Chinese Academy of Sciences (CAS), Beijing 100190, China}

\author{Rong-Gen Cai}
\email{caironggen@nbu.edu.cn}
\affiliation{Institute of Fundamental Physics and 
Quantum Technology, Ningbo University, Ningbo, 315211, China}

\author{Shao-Jiang Wang}
\email{schwang@itp.ac.cn (corresponding author)}
\affiliation{Institute of Theoretical Physics, Chinese Academy of Sciences (CAS), Beijing 100190, China}
\affiliation{Asia Pacific Center for Theoretical Physics (APCTP), Pohang 37673, Korea}

\begin{abstract}
Recently, a $3\sim4\sigma$ preference for dynamical dark energy has been reported by the Dark Energy Spectroscopic Instrument (DESI) collaboration, which has inspired hot debates on new physics or systematics. In this paper, we reveal that this preference is significantly biased by an external low-redshift supernova (low-$z$ SN) sample, which was combined with the Dark Energy Survey SN program (DES-SN) in their Year-Five data release (DESY5). Using the intercept in the SN magnitude-distance relation as a diagnostic for systematics, we find not only large dispersions but also a large discrepancy in the low-$z$ SN sample when compared to the high-$z$ DES-SN sample within the single DESY5 compilation, in contrast to the uniform behavior found in the PantheonPlus data. Correcting for this low-$z$ systematics with or without including the cosmic microwave background data can largely reduce the preference for dynamical DE to be less than $2\sigma$. Therefore, the DESI preference for dynamical DE is biased by some unknown systematics in the low-$z$ SN sample. 
\end{abstract}
\maketitle

\section{Introduction} \label{sec:intro}

The standard Lambda-cold-dark-matter ($\Lambda$CDM) model, consisting of radiation, baryonic matter, cold dark matter, and a cosmological constant $\Lambda$ in a spatially flat, isotropic and homogeneous Universe with primordial perturbations provided by a nearly scale-invariant adiabatic Gaussian random scalar field, has passed a wide variety of cosmological tests~\cite{SupernovaSearchTeam:1998fmf,SDSS:2005xqv,SPT:2019qkp,Planck:2018vyg,eBOSS:2020yzd,DES:2021wwk,ACT:2023kun,Bennett:1996ce,WMAP:2012nax}. Despite its great success over past two decades, a few tensions~\cite{Perivolaropoulos:2021jda} including the $\sim 5\sigma$ Hubble tension~\cite{Bernal:2016gxb,Verde:2019ivm,Knox:2019rjx,Riess:2020sih,Freedman:2021ahq}, a $\sim3\sigma$ $S_8$ tension~\cite{DiValentino:2020vvd,Abdalla:2022yfr}, and a $\sim4\sigma$ $\gamma$ tension~\cite{Nguyen:2023fip} as well as a $\sim3\sigma$ $\delta H_0$ tension~\cite{Yu:2022wvg} seem to breakdown this concordant model. In particular, whether the cosmological constant $\Lambda$ should be replaced by a dynamical dark energy (DE) to drive the late-time acceleration would call for special cautions.  

As one of the stage-III surveys, the Dark Energy Survey (DES) explores the properties of dark matter and DE with unprecedented precision and accuracy through multiple probes like Type Ia supernovae (SNe Ia), weak lensings, large-scale structures, galaxy clusters~\cite{DES:2005dhi,Bernstein:2011zf,DES:2016jjg}. As the most updated Stage IV survey, the recent release of the Dark Energy Spectroscopic Instrument (DESI) for baryon acoustic oscillations (BAO)~\cite{DESI:2024mwx,DESI:2024uvr,DESI:2024lzq} together with full 5-year photometrically classified 1635 SNe Ia (DES-SN), after combined with previous external 194 low-$z$ SNe Ia (low-$z$ SN) in the DES Year-Five Data Release (DESY5)~\cite{DES:2024jxu,DES:2024hip}, have triggered heated debates on the existence of dynamical DE. Specifically, the joint constraint of cosmic microwave background (CMB) Planck data, DESI DR1 (Y1 BAO measurements), and DESY5 (low-$z$ SN+DES-SN) compilation surprisingly favors a dynamical DE with phantom crossing, showing a $3.9\sigma$ tension with the $\Lambda$CDM model~\cite{DESI:2024mwx,DESI:2024hhd}. A similar but weaker preference ($\sim 2\sigma$)~\cite{Park:2024vrw} of dynamic DE is also achieved without the participation of DESI DR1 and DESY5. These hints for new physics have attracted significant interests in the community with subsequent numerous theoretical studies on different kinds of dynamical DE models, e.g. Refs.~\cite{Tada:2024znt,Berghaus:2024kra,Shlivko:2024llw,Ramadan:2024kmn,Bhattacharya:2024hep,Giare:2024smz,Allali:2024cji,Croker:2024jfg,Yang:2024kdo,Escamilla-Rivera:2024sae,Yin:2024hba,Gu:2024jhl,Wang:2024qan,Cortes:2024lgw,Colgain:2024xqj,Carloni:2024zpl,DESI:2024kob,DESI:2024aqx,Luongo:2024fww,Mukherjee:2024ryz,Dinda:2024kjf,DES:2024fdw,Li:2024qus,Gao:2024ily,Fikri:2024klc,Li:2025owk}.

In parallel with theoretical explanations, inconsistencies between recent new measurements (DESI DR1, DESY5) and previous high-quality measurements, such as SDSS BAO~\cite{BOSS:2016wmc,eBOSS:2020yzd} and PantheonPlus SNe~\cite{Riess:2021jrx,Brout:2022vxf,Scolnic:2021amr,Brout:2021mpj,Peterson:2021hel,Carr:2021lcj,Popovic:2021yuo}, have also been widely debated~\cite{Chan-GyungPark:2024brx,Chan-GyungPark:2025cri,Liu:2024gfy,Colgain:2024ksa,Colgain:2024mtg,Wang:2024pui,Huang:2024qno,Dhawan:2024gqy,RoyChoudhury:2024wri,Gialamas:2024lyw,Efstathiou:2024xcq,Notari:2024zmi,Sakr:2025daj}. For example, the combination of DESI LRG1 and LRG2 data is found to be the key driver for the preference of dynamical DE~\cite{Liu:2024gfy,Wang:2024pui,Huang:2024qno,DESI:2024mwx,Colgain:2024xqj}. Nevertheless, these two data exhibit mild tension with SDSS BAO~\cite{eBOSS:2020yzd} for unknown reasons. What's more, by comparing the mean magnitude-distance residuals between PantheonPlus and DESY5, Ref.~\cite{Efstathiou:2024xcq} found a $\sim0.04-0.05$ magnitude (mag.) difference in the overlapping regime, indicating unidentified SNe systematics in DESY5 (see, however, Ref.~\cite{DES:2025tir}). Correcting this offset would reduce the preference for dynamical DE, bringing cosmological constraints of the corrected DESY5 back to $\Lambda$CDM. Therefore, it cannot simply rule out the possibility of potential systematics in DESY5.

The low-$z$ SN sample consists of 194 SNe Ia, including 8 SNe Ia from the Carnegie Supernova Project (CSP)~\cite{Krisciunas:2017yoe}, 68 SNe Ia from the Center for Astrophysics (CfA)~\cite{Hicken:2009df,Hicken:2012zr}, and 118 SNe Ia from the PanSTARRS Foundation Supernova Survey (Foundation)~\cite{Foley:2017zdq}. On the other hand, the PantheonPlus sample consists of 741 SNe Ia at low redshift $z<0.1$, 184 of which are in common with the low-$z$ SN sample of DESY5 compilation, that is, 118 SNe from Foundation, 59 SNe from CfA, and 7 SNe from CSP samples. Including the low-$z$ SN sample in the DESY5 compilation with the least systematics control in cosmological analyses is controversial in practice. As highlighted in Refs.~\cite{DES:2024jxu,DES:2024hip}, the spectroscopically confirmed low-$z$ SN sample (different from the photometrically classified DES-SN in the observational method) is the most critical source of systematics due to a lack of homogeneity and complete calibrations. Specifically, the nuisance parameters $\alpha, \beta$ used to correct for SN color and stretch dependencies show significant discrepancies between low-$z$ SN and DES-SN samples. Given that the properties and selection effects of the low-$z$ sample still lack sufficient understanding, additional systematics checks are necessary.

In this paper, we first cross-compare DESI DR1 and DESY5 SNe with the previous high-quality SDSS eBOSS BAO and PantheonPlus SNe, and confirm that the key driver of dynamical DE is the additional external low-$z$ SN sample in forming the DESY5 compilation. We then propose to diagnose the systematics of DESY5 and quantify its preference for dynamic DE through the dubbed $a_B$ consistency developed in Refs.~\cite{Huang:2024erq,Huang:2024gfw}. By performing simple corrections to DESY5, we find the fully de-biased DESY5, after combining with DESI DR1/DR2 regardless of CMB, shows no significant deviation from the $\Lambda$CDM model. Therefore, it is too early to claim dynamical DE.

\section{Data and method}

\begin{figure*}
\centering
\includegraphics[width=0.49\textwidth]{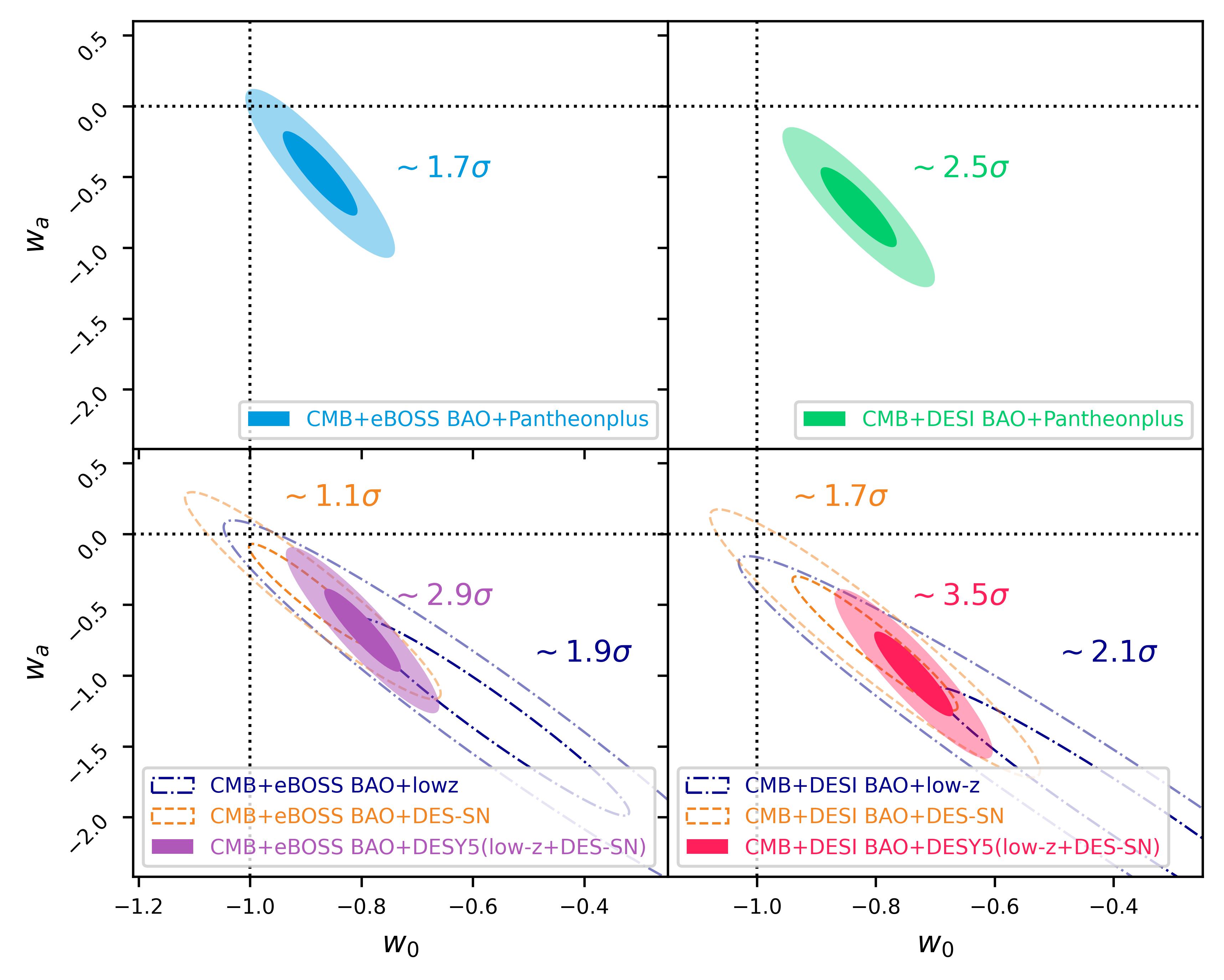}
\includegraphics[width=0.49\textwidth]{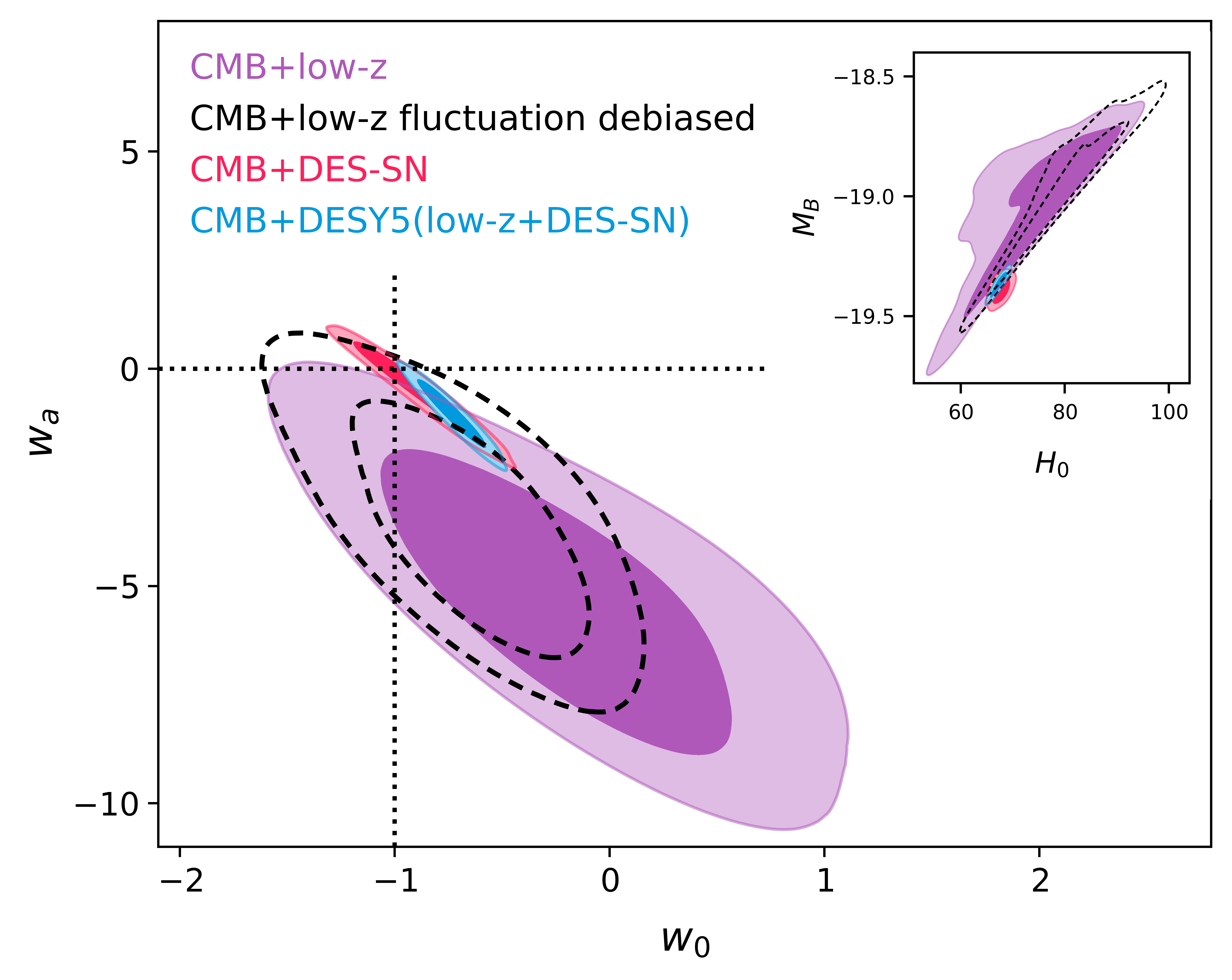}\\
\caption{Left: The $w_0-w_a$ constraits from CMB (Planck 2018) + BAO (eBOSS DR16/DESI DR1) + SNe (PantheonPlus/DESY5). Right: The $w_0-w_a$ constraints from CMB+DESY5/DES-SN/low-$z$ (with and without fluctuations debiased).}
\label{fig:w0wacomparison}
\end{figure*}

As indicated in Refs.~\cite{DESI:2024mwx,DESI:2024hhd}, different late-time observations seem to have different levels of support for the dynamical DE. To diagnose which dataset contributes the most to the preference for dynamical DE, we take all the high-quality DESI DR1~\cite{DESI:2024mwx}, DESY5 (low-$z$ SN + DES-SN)~\cite{DES:2024jxu,DES:2024hip}, SDSS eBOSS BAO~\cite{eBOSS:2020yzd}, PantheonPlus~\cite{Riess:2021jrx,Brout:2022vxf,Scolnic:2021amr,Brout:2021mpj,Peterson:2021hel,Carr:2021lcj,Popovic:2021yuo}, and Planck CMB~\cite{Planck:2018vyg} into considerations for completed comparisons. The UNION3 compilation~\cite{Rubin:2023ovl} is not included in this study as they have not released a Hubble diagram yet.

Similar to PantheonPlus SNe analysis, we construct the distance moduli residual vector of DESY5 SNe as $\Delta \vec{D} =m_B^\mathrm{std}-M_B-\mu_{\mathrm{model} }$, where standardized apparent magnitudes $m_B^\mathrm{std}$ have already corrected for stretch, color, simulation bias (selection bias, Malmquist bias, and light curve fitting bias), and mass step effects. The likelihood is defined as $-2\ln(\mathcal{L} )=\Delta \vec{D^{T}}C^{-1}_{\mathrm{stat+syst} }\Delta\vec{D}$. We adopt the commonly used built-in Planck CMB (plik TTTEEE + lowl + lowE + CMB lensing), SDSS eBOSS BAO (eight BAO-only data points), and PantheonPlus (local SNe at $z<0.01$ are discarded) in MontePython~\cite{Audren:2012wb}, along with extra self-written DESI BAO and DESY5 likelihoods. In what follows, we perform the Monte Carlo Markov Chain (MCMC) on the $w_0w_a$CDM model with Planck CMB + arbitrarily different combinations of BAO and SNe to locate the possible systematics direction.

\section{Analysis}

\subsection{Systematics identification}

\begin{figure*}
\centering
\includegraphics[width=0.9\textwidth]{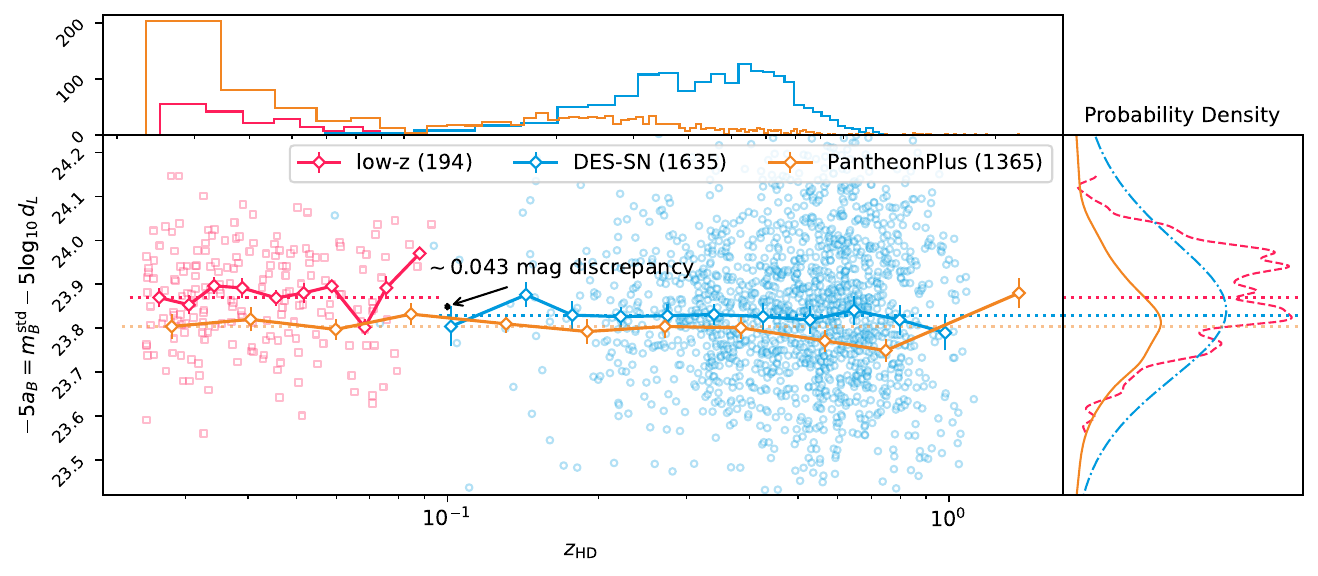}\\
\includegraphics[width=0.9\textwidth]{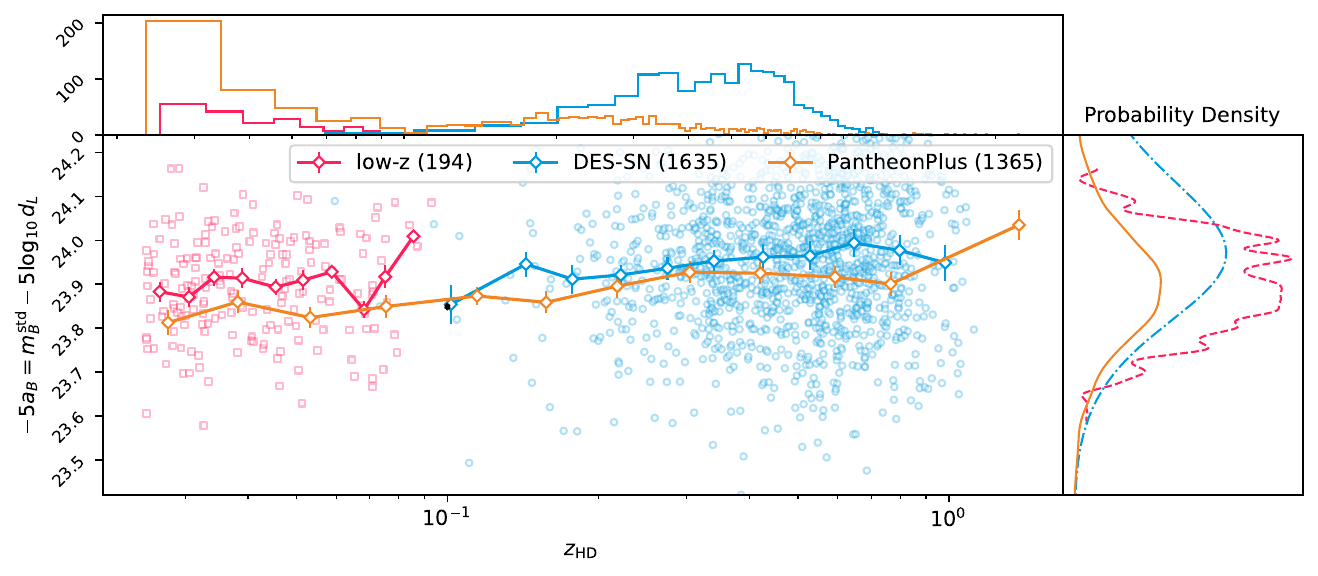}\\
\caption{\textit{Top:} The $a_B$ diagnosis: the distributions of $-5a_B$ for low-$z$, DES-SN, and PantheonPlus SNe (only 1365 SNe with $z\ge 0.0233$ are considered). The fiducial cosmology is assumed from the best-fit Planck $\Lambda$CDM with $\Omega_m=0.315$. Note that the discrepancy in $-5a_B$ between low-$z$ and DES-SN samples is independent of fiducial $\Omega_m$ values. The top panel is the redshift distribution for the number of SNe while the right panel is the probability density distributions of $-5a_B$ for different SN samples. \textit{Bottom:} The fiducial cosmology used to calculate $-5a_B$ is changed into the best-fit $w_0w_a$CDM model from DESI DR2~\cite{DESI:2025zgx} with $\Omega_m=0.352_{-0.018}^{+0.041}$, $w_0=-0.48_{-0.17}^{+0.35}$, and $w_a<-1.34$.}
\label{fig:intercept}
\end{figure*}

In the Appendix, we present all constraints of the $w_0w_a$CDM model of this study in Tab.~\ref{tab:w0waCDMcomparison} and Tab.~\ref{tab:w0waCDMcorrection}. The Tab.~\ref{tab:w0waCDMcomparison} is visualized in the left panel Fig.~\ref{fig:w0wacomparison} for their marginalized $w_0-w_a$ distributions, where the base preference for dynamical DE from Planck-CMB+eBOSS-BAO+PantheonPlus is $\sim1.7\sigma$ (blue contour in the upper left), not significant yet. Either replacing the eBOSS BAO with DESI DR1 (green contour in the upper right) or the PantheonPlus SNe with DESY5 SNe (purple contour in the bottom left) would increase preferences to $\sim2.5\sigma$ and $\sim2.9\sigma$, respectively, but still not decisive to claim dynamical DE. In particular, when getting rid of the low-$z$ SN sample, the significance drops from $\sim2.9\sigma$ to insignificant $\sim1.1\sigma$ (orange contour in the bottom left). Even though CMB+DESI DR1+DESY5 after both of the above replacements presents a more significant $\sim3.5\sigma$ preference (red contour in the bottom right), getting rid of the low-$z$ SN sample would once again drop the significance from $\sim3.5\sigma$ back to $\sim1.7\sigma$ (orange dashed contour in the bottom right). On the contrary, both cases using low-$z$ SNe (blue dot-dashed contours in the bottom row) prefer a deviation from the $\Lambda$CDM model with a direction towards $w_0>-1$ and $w_a<0$. Therefore, a minor inconsistency emerges between low-$z$ SNe and DES-SN samples.

To further separate the interference from different BAO data and reflect the underlying preference for dynamical DE from different SNe samples, we can solely use Planck CMB to calibrate low-$z$, DES-SN, and their combined (DESY5), respectively. The marginalized $w_0-w_a$ and $H_0-M_B$ distributions are shown in the right panel of Fig.~\ref{fig:w0wacomparison}. The purple $w_0-w_a$ contour of low-$z$ SN samples shows a significant deviation and thus dominates the preference for dynamical DE in the blue DESY5 contour since the red DES-SN contour alone coincides perfectly with $\Lambda$CDM ($\sim0.6\sigma$). However, this low-$z$-dominated deviation is likely to be a mirage since its $H_0-M_B$ contour in the inset shows a rather weak constraint and irregular marginalized distribution, suggesting the multi-sample mixed low-$z$ SNe does not have a consistent absolute magnitude $M_B$. Hence, there are substantial systematics in the low-$z$ SNe sample, which can be corrected with a method~\eqref{eq:flucdebias} detailed below in the next two subsections, as shown with black dashed contours in both Fig.~\ref{fig:w0wacomparison} and its inset.

\subsection{Systematics diagnosis}

\begin{figure*}
\centering
\includegraphics[width=0.45\textwidth]{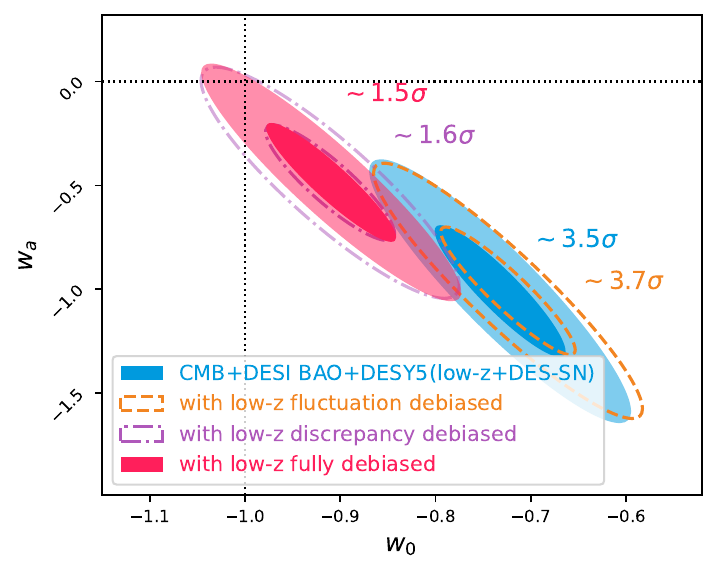}
\includegraphics[width=0.45\textwidth]{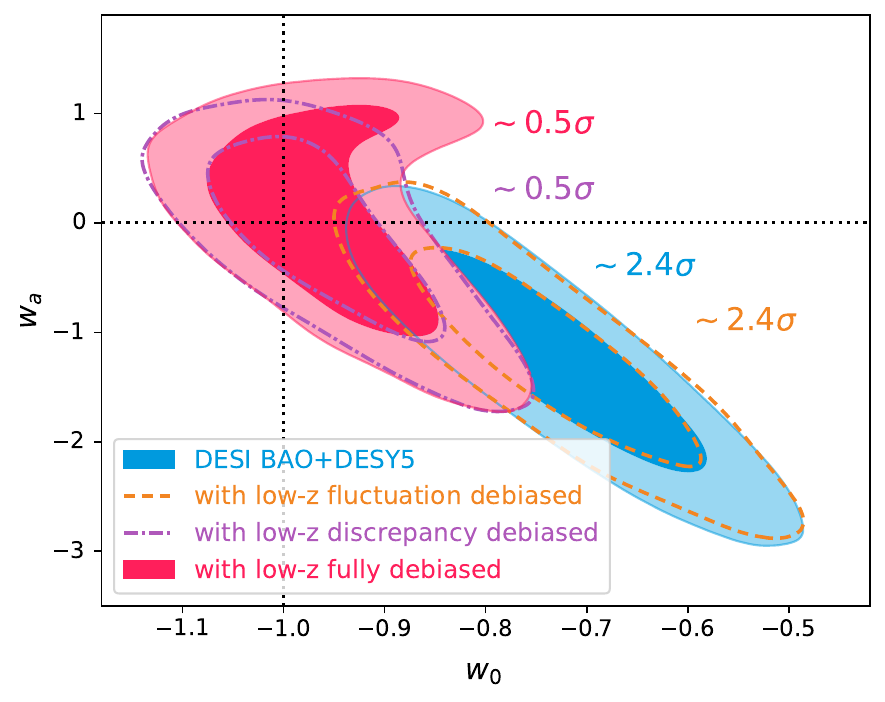}
\includegraphics[width=0.45\textwidth]{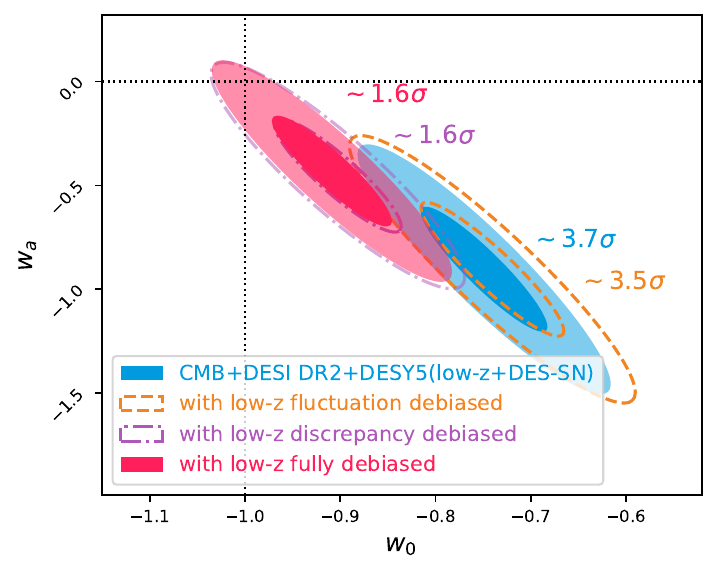}
\includegraphics[width=0.45\textwidth]{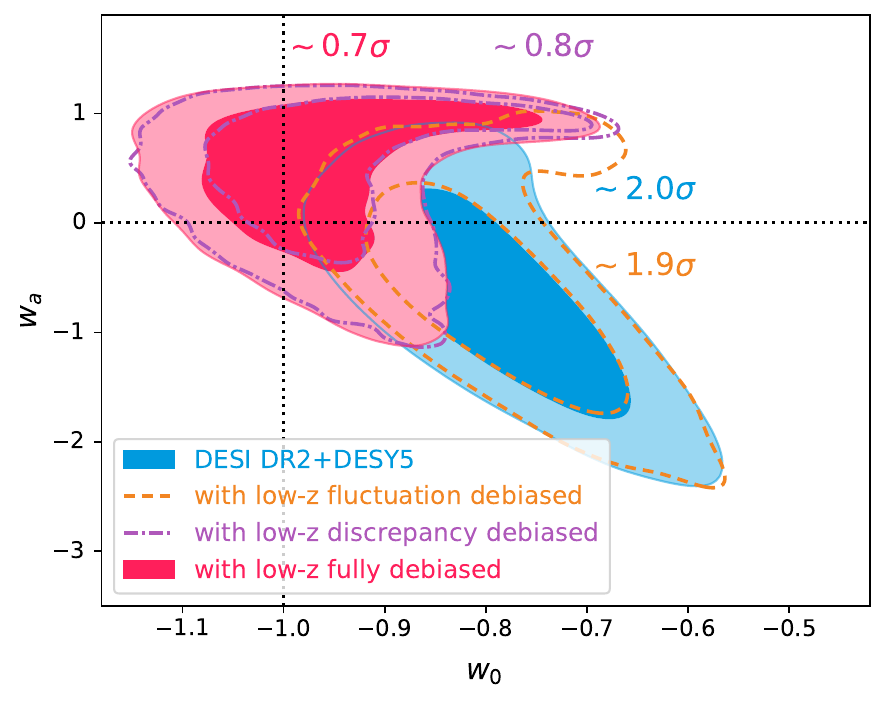}\\
\caption{Left: The marginalized contours on parameters $w_0$ and $w_a$ for Planck CMB+DESI BAO+DESY5 with different de-biasing corrections. Right: the same as the left panel but excluding Planck-CMB with a fiducial $H_0=67.4$ km/s/Mpc. The DESI DR1 (Y1 BAO) and DR2 (Y3 BAO) are used in the top and bottom rows, respectively.}
\label{fig:lowzcomparison}
\end{figure*}

To diagnose systematics for low-$z$ SNe, we rebuild the magnitude-(logarithmic) distance relation
\begin{align}\label{eq:aBdefinition}
m_{B,i}^\mathrm{std}&=5\lg\hat{d}_L(z_i)-5a_B,\\
-5a_B&\equiv M_B+5\lg\frac{c/H_0}{\mathrm{Mpc}}+25
\end{align}
for both low-$z$ and DES-SN samples, where the dimensionless luminosity distance $\hat{d}_L(z_i)=D_L(z)/(c/H_0)=(1+z_\mathrm{hel})\int_0^{z_\mathrm{HD}}\mathrm{d}z'/E(z')$ is computed from the best-fit Planck $\Lambda$CDM fiducial model for the reduced Hubble parameter $E(z)^2\equiv H(z)^2/H_0^2=\Omega_m(1+z)^3+(1-\Omega_m)$. Recall that all light-curve model corrections are used to standardize the apparent magnitude $m_{B,i}^\mathrm{std}$, and hence the intercept $-5a_B$ as a degeneracy parameter between $H_0$ and $M_B$ is expected to be a stable value for a single SNe compilation. The gist of $a_B$ diagnosis is to check whether the intercept $-5a_B$ (a systematics-sensitive and model-sensitive quantity) is consistent among different SNe samples. As any observational systematics in measuring $m_{B,i}^\mathrm{std}$ or theoretical systematics in modeling $5\lg \hat{d}_L(z_i)$ induced by either uncounted peculiar velocity effect on the redshift $z_i$ measurement or even new physics modeling in $\hat{d}_L(z)$ would cause a deviation in the intercept $-5a_{B,i}=m_{B,i}^\mathrm{std}-5\lg\hat{d}_L(z_i)$, we can quickly locate the redshift ranges of either systematics or new physics from redshift distributions of $-5a_B$.

Detailed $a_B$ diagnosis is visualized in the top panel of Fig.~\ref{fig:intercept} with a fiducial Planck-$\Lambda$CDM cosmology. The high-quality DES-SN from a single telescope has the most stable $a_B$ distributions among its spanned redshift bins, perfectly agreeing with the Planck $\Lambda$CDM cosmology assumed in the top panel of Fig.~\ref{fig:intercept}. By contrast, the multi-sample mixed low-$z$ SNe shows irregular fluctuations in $-5a_B$, and its weighted average is $\sim0.043$ mag. higher than that of DES-SN, which is qualitatively similar to the previous study by~\cite{Efstathiou:2024xcq} but with the $\sim0.04$ mag. offset estimated on $\Delta m_\mathrm{offset}:=(m_\mathrm{Pantheon+}^{\mathrm{low}-z}-m_\mathrm{DESY5}^{\mathrm{low}-z})-( m_\mathrm{Pantheon+}^{\mathrm{high}-z}-m_\mathrm{DESY5}^{\mathrm{high}-z})\approx(-0.05)-(-0.01)=-0.04$ mag. between PantheonPlus and DESY5 compilations in their low- and high-redshift ranges after assuming a constant $M_B$. As a response, Ref.~\cite{DES:2025tir} has explained partially this offset with improvements to the intrinsic scatter model and mass step estimate by reducing the offset from $\sim0.04$ mag. to $\Delta\mu_\mathrm{offset}:=(\mu_\mathrm{Pantheon+}^{\mathrm{low}-z}-\mu_\mathrm{DESY5}^{\mathrm{low}-z})-(\mu_\mathrm{Pantheon+}^{\mathrm{high}-z}-\mu_\mathrm{DESY5}^{\mathrm{high}-z})=(\mu_\mathrm{Pantheon+}^{\mathrm{low}-z}-\mu_\mathrm{Pantheon+}^{\mathrm{high}-z})-(\mu_\mathrm{DESY5}^{\mathrm{low}-z}-\mu_\mathrm{DESY5}^{\mathrm{high}-z})=0.019-0.043=-0.024$ mag., which can be further reduced to $-0.008$ mag. after considering a $\sim0.016$ mag. difference in selection functions as elaborated in Table 1 of Ref.~\cite{DES:2025tir}. However, such a negligible $-0.008$ mag. offset in $\Delta\mu_\mathrm{offset}$ is actually achieved between two different SN compilations, leaving an equivalent $0.019+0.008\sim0.027$ mag. difference between low-$z$ and high-$z$ SNe within the single DESY5 compilation, which is still apparent in the difference of $-5a_B$ between low-$z$ and high-$z$ parts of the single DESY5 compilation. Therefore, the key difference of our current study from the previous two analyses~\cite{Efstathiou:2024xcq,DES:2025tir} is that they considered the low/high-redshift difference among two different SN compilations (PantheonPlus and DESY5) while we focus on the low/high-redshift difference within the same SN compilation.

This $\sim0.043$ mag. discrepancy in $-5a_B$ within the single DESY5 compilation appears to emerge suddenly around $z\sim0.1$ where the Universe is still homogeneous, hence we can safely rule out the local inhomogeneous effects. This sudden transition also cannot originate from homogeneous-scale new physics since the PantheonPlus compilation, containing a larger number of low-$z$ well-calibrated SNe, does not admit any such low-$z$ transition. If a homogeneous-scale transition model could compensate for this $a_B$ discrepancy within the DESY5 compilation, the same model would, in turn, jeopardize the original $a_B$ consistency of PantheonPlus and reproduce a new $a_B$ discrepancy within the PantheonPlus compilation. \textit{This is the key argument of this paper}. To explicitly show this, we illustrate in the bottom panel of Fig.~\ref{fig:intercept} with the fiducial cosmology replaced by the best-fit $w_0w_a$CDM model from DESI DR2~\cite{DESI:2025zgx}. In this case, not only the low-$z$ and high-$z$ parts of PantheonPlus sample admit different $-5a_B$ values, but also the whole intercept $-5_B$ is no longer a constant anymore but slightly tilted for both DESY5 and PantheonPlus compilations.  In addition, the 1D probability density distribution (red dashed line) of low-$z$ intercept $-5a_B$ in Fig.~\ref{fig:intercept} is non-Gaussian, unlike the other two cases, suggesting the low-$z$ SNe suffer from large scatters and do not follow a consistent magnitude-distance relation. As a comparison, the PantheonPlus SNe contain even more subsamples (18 samples) but still converge to a Gaussian distribution around the weighted average, indicating their well-corrected SNe have followed an accordant magnitude-distance relation just as the DES-SN does. Therefore, the strange transition and large scatter in the low-$z$ intercept $-5a_B$ can only be attributed to systematics (including an $M_B$ transition). Regardless of its origin, these low-$z$ SNe can only be used for cosmological analyses after correcting this mismatch intercept.

\subsection{Systematics corrections}

The low-$z$ SNe systematics call for better sample selections and calibrations, which goes well beyond our current study. Based on the $a_B$ diagnosis, we propose a simple but rough correction procedure to the low-$z$ sample, which is, however, sufficient for our purpose to re-examine the preference for dynamic DE. Since the intercept $-5a_B$ measured from high-$z$ SNe is less affected by systematics from the peculiar velocity effect than that measured from low-$z$ SNe, the systematics correction for low-$z$ SNe is done by matching the low-$z$ intercept to the high-$z$ intercept. Specifically, our correction procedures include three kinds: 
(i) Smoothing the irregular fluctuations in the intercept $-5a_B$ for each redshift bin of low-$z$ SNe to match the weighted average of all low-$z$ SNe,
\begin{align}\label{eq:flucdebias}
m_{B,i\in\mathrm{low}-z}^\mathrm{fluc-debias}:=5\lg\hat{d}_L(z_{i\in\mathrm{low}-z})-5\overline{a_{B,\mathrm{low}-z}}\,;
\end{align}
(ii) Reducing the low-$z$ $m_B$ by $0.043$ magnitude so that the weighted average $-5a_B$ of low-$z$ SNe also matches the weighted average of DES-SN,
\begin{align}\label{eq:dispdebias}
m_{B,i\in\mathrm{low}-z}^\mathrm{disp-debias}&:=5\lg\hat{d}_L(z_{i\in\mathrm{low}-z})-5a_{B,i\in\mathrm{low}-z}-0.043\nonumber\\
&\equiv m_{B,i\in\mathrm{low}-z}^\mathrm{std}-0.043\,;
\end{align}
(iii) Combining correction processes (i) and (ii) together to have a fully de-biased low-$z$ SN sample,
\begin{align}\label{eq:fullydebias}
m_{B,i\in\mathrm{low}-z}^\mathrm{fully-debias}&:=5\lg\hat{d}_L(z_{i\in\mathrm{low}-z})-5\overline{a_{B,\mathrm{low}-z}}-0.043\nonumber\\
&\equiv5\lg\hat{d}_L(z_{i\in\mathrm{low}-z})-5\overline{a_{B,\mathrm{DES-SN}}}.
\end{align}
It is worth noting that debiasing the $a_B$ fluctuation with~\eqref{eq:flucdebias} for low-$z$ sample would slightly change their correlations to DES-SN, so does the fully debiased low-$z$ sample with~\eqref{eq:fullydebias}. However, this would not affect our conclusion since debiasing the $a_B$ discrepancy with~\eqref{eq:dispdebias} alone has already diminish the dark energy significance as shown shortly below, and shifting the apparent magnitudes of low-$z$ SNe by the same amount $0.043$ in~\eqref{eq:dispdebias} would not change their correlations to DES-SN.

As a preliminary test, we find a $\sim2.2\sigma$ preference for dynamical DE from CMB+original low-$z$ SNe decreases to $\sim1.5\sigma$ after roughly de-biasing the fluctuations in $-5a_B$ as shown in the first group of constraints in Tab.~\ref{tab:w0waCDMcorrection}, which also slightly regularizes the $H_0-M_B$ contour (black unfilled contours) in the inset of the right panel of Fig.~\ref{fig:w0wacomparison}. We further test our correction procedures when including DESI Y1/Y3 BAO as shown in the upper/lower-left panels of Fig.~\ref{fig:lowzcomparison}, respectively, as well as the second/fourth groups of constraints in Tab.~\ref{tab:w0waCDMcorrection}. In the upper/lower-left panel of Fig.~\ref{fig:lowzcomparison} with inclusions of DESI Y1/Y3 BAO, correcting for the fluctuations in $-5a_B$ alone can cause little deviation in the $w_0-w_a$ contour, however, solely correcting for the low-$z$ discrepancy in $-5_B$ without debiasing the fluctuations first is already sufficient to reduce the preference for dynamical DE from $\sim3.5\sigma/3.7\sigma$ to $\sim1.6\sigma$, almost overlapping with the fully de-biased correction case with $\sim1.5\sigma/1.6\sigma$. A larger reduction to $\sim0.5\sigma/0.7\sigma$ and the same pattern also emerges in the upper/lower-right panels of Fig.~\ref{fig:lowzcomparison} with DESI Y1/Y3 BAO when further excluding Planck-CMB assuming fiducial $H_0=67.4$ km/s/Mpc as shown in the third and fifth groups of constraints in Tab.~\ref{tab:w0waCDMcorrection}. A different $H_0$ prior only changes the degenerated constraints with respect to the sound horizon $r_d$ and absolute magnitude $M_B$ but still preserves the same reduced tension in the $w_0-w_a$ plane for this purely late-Universe data combination (DESI BAO+DESY5) without early-Universe Planck-CMB calibration. Therefore, it is the $\sim0.043$ mag. discrepancy in $-5a_B$ between low-$z$ and DES-SN that plays a leading role in driving the preference for dynamical DE.

\section{Conclusions and Discussions}

Recent cosmological analyses from DESI DR1 and DR2 have attracted wide interest but also raised significant challenges and hot debate to the standard $\Lambda$CDM. In this paper, we reduce the challenge by locating some systematics at low-redshift supernovae. The multi-sample mixed low-$z$ SN sample in DESY5 compilation, suffering from large scatters, fails to rebuild a consistent intercept of the magnitude-distance relation and admits a $\sim0.043$ magnitude discrepancy with respect to the DES-SN at higher redshifts, which largely drives the preference for dynamical DE. After the first and second versions of our preprint, the DESI DR2 paper~\cite{DESI:2025zgx} has explicitly tested this observation by directly getting rid of low-$z$ SNe in CMB+DESI DR2+DESY5($z>0.1$), where the preference for dynamical DE is reduced to be $2\sigma$ as shown in the middle panel of Fig.~14. After roughly correcting this discrepancy in the intercept for the low-$z$ SN samples, we find the corrected datasets (Planck+DESI DR1/DR2+DESY5) favor dynamical DE only at $1.5\sigma/1.6\sigma$ levels, not significant at all to call for dynamical DE. Therefore, the success of the standard $\Lambda$CDM has not been overturned.

Although we have identified the systematics involving $a_B$ inconsistency in the low-$z$ SN sample within the single DESY5 compilation is largely responsible for the immature claim of dynamical DE, the DESI DR1 BAO data alone indeed shows some deviations from the Planck-$\Lambda$CDM model at higher redshifts, especially for those luminous red galaxies (LRG) between $0.6<z<0.8$~\cite{DESI:2024mwx}, which have also been considered recently in Refs.~\cite{Wang:2024pui,Liu:2024gfy} as the possible drives for the preference of dynamical DE. This SN-independent deviation from Planck-$\Lambda$CDM model has been re-enforced recently in the DESI DR2 papers~\cite{DESI:2025zgx,DESI:2025zpo,DESI:2025fii,DESI:2025ejh} with an explicit discrepancy in $\Omega_m$ between Planck-CMB and DESI-BAO and a clear deviation from Planck-$\Lambda$CDM model for LRG2 BAO measurements. Whether this DESI-Planck tension should be better explained by dynamical DE (with phantom crossing) is optional~\cite{Guo:2004fq,Feng:2004ad,Feng:2004ff}, as it can be equally accounted for by some modified gravity with non-minimal coupling~\cite{Ye:2024ywg,Pan:2025psn,Wolf:2025jed}, mimicking thawing DE~\cite{Chen:2024vuf,Wolf:2024stt,Wolf:2024eph} even with phantom crossing~\cite{Cai:2021wgv,Chakraborty:2025syu,Khoury:2025txd}.

\begin{acknowledgments}
We thank Tamara Davis, George Efstathiou, Bin Hu, Yipeng Jing, Cheng Zhao, Gong-Bo Zhao, Jian-Qi Liu, and Yan-Hong Yao for helpful discussions. 
This work is supported by 
the National Key Research and Development Program of China Grants No. 2021YFA0718304, No. 2021YFC2203004, and No. 2020YFC2201501,
the National Natural Science Foundation of China Grants No. 12422502, No. 12105344, No. 11821505, No. 12235019, No. 11991052, No. 12447101, No. 12073088, and No. 11947302,
the China Manned Space Program with Grant No. CMS-CSST-2025-A01,
and the Postdoctoral Fellowship Program of CPSF.
We also acknowledge the use of the HPC Cluster of ITP-CAS.
\end{acknowledgments}

\appendix

\begin{table*}[!htbp]
\begin{flushleft}
\caption{Constraints on $w_0w_a$CDM model with the significance level of the tension given with respect to the $\Lambda$CDM model.}
\label{tab:w0waCDMcomparison}
\renewcommand{\arraystretch}{1.2}
\begin{tabular}{|ccccccc|}
\hline
\multicolumn{1}{|c|}{Datasets} & \multicolumn{1}{c|}{$H_0$} & \multicolumn{1}{c|}{$M_B$} & \multicolumn{1}{c|}{$\Omega_m$} & \multicolumn{1}{c|}{$w_0$} &  \multicolumn{1}{c|}{$w_a$} & Tensions\\ 
\hline

\multicolumn{7}{|c|}{Planck CMB + }                      \\ \hline
\multicolumn{1}{|c|}{low-$z$} & 
\multicolumn{1}{c|}{$74.0\pm 8.0$} &
\multicolumn{1}{c|}{$-19.05^{+0.26}_{-0.17}$} &
\multicolumn{1}{c|}{$0.246^{+0.030}_{-0.066}$} & \multicolumn{1}{c|}{$-0.20\pm0.50$} & 
\multicolumn{1}{c|}{$-5.4\pm2.0$} & 
$\sim 2.2\sigma$ \\
\hline

\multicolumn{1}{|c|}{DES-SN} & 
\multicolumn{1}{c|}{$67.7\pm1.2$} & 
\multicolumn{1}{c|}{$-19.390\pm0.037$} & 
\multicolumn{1}{c|}{$0.312^{+0.011}_{-0.012}$} & \multicolumn{1}{c|}{$-0.90^{+0.16}_{-0.20}$ }& \multicolumn{1}{c|}{$-0.44^{+0.79}_{-0.56}$}  & $\sim 0.6\sigma$ \\
\hline

\multicolumn{1}{|c|}{DESY5(low-$z$+DES-SN)} & 
\multicolumn{1}{c|}{$67.4^{+1.1}_{-1.0}$} & 
\multicolumn{1}{c|}{$-19.367^{+0.038}_{-0.031}$} &
\multicolumn{1}{c|}{$0.315^{+0.010}_{-0.012}$} & \multicolumn{1}{c|}{$-0.73\pm0.10$} & 
\multicolumn{1}{c|}{$-1.05\pm0.51$}  & 
$\sim 2.4\sigma$ \\ 
\hline

\multicolumn{7}{|c|}{Planck CMB + DESI BAO DR1 +}\\
\hline

\multicolumn{1}{|c|}{low-$z$} & 
\multicolumn{1}{c|}{$64.1\pm3.0$} & 
\multicolumn{1}{c|}{$-19.450\pm0.089$} &\multicolumn{1}{c|}{$0.351^{+0.029}_{-0.036}$} & \multicolumn{1}{c|}{$-0.37^{+0.28}_{-0.35}$} & \multicolumn{1}{c|}{$-2.00^{+1.1}_{-0.75}$} & $\sim 2.1\sigma$ \\ \hline
\multicolumn{1}{|c|}{DES-SN} & 
\multicolumn{1}{c|}{$67.6\pm 1.3$} &
\multicolumn{1}{c|}{$-19.373\pm 0.018$} &
\multicolumn{1}{c|}{$0.313^{+0.010}_{-0.012}$} & \multicolumn{1}{c|}{$-0.79\pm0.14$} & \multicolumn{1}{c|}{$-0.83^{+0.49}_{-0.40}$}  & 
$\sim 1.7\sigma$ \\ \hline
\multicolumn{1}{|c|}{DESY5(low-$z$+DES-SN)} & 
\multicolumn{1}{c|}{$67.25\pm0.72$} & 
\multicolumn{1}{c|}{$-19.370\pm0.017$} &
\multicolumn{1}{c|}{$0.316\pm0.0077$} & \multicolumn{1}{c|}{$-0.73\pm0.069$} & 
\multicolumn{1}{c|}{$-1.01\pm0.32$}  & 
$\sim 3.5\sigma$\\
\hline

\multicolumn{7}{|c|}{Planck CMB + eBOSS BAO DR16+ } \\
\hline 

\multicolumn{1}{|c|}{low-$z$} & 
\multicolumn{1}{c|}{$64.5^{+2.0}_{-2.4}$} & 
\multicolumn{1}{c|}{$-19.443^{+0.059}_{-0.070}$} &
\multicolumn{1}{c|}{$0.345\pm0.024$} & 
\multicolumn{1}{c|}{$-0.56\pm0.24$} & 
\multicolumn{1}{c|}{$-1.3^{+0.73}_{-0.65}$}  & 
$\sim 1.9\sigma$ \\
\hline

\multicolumn{1}{|c|}{DES-SN} & 
\multicolumn{1}{c|}{$67.4\pm1.0$} & 
\multicolumn{1}{c|}{$-19.393\pm0.016$} &
\multicolumn{1}{c|}{$0.315\pm0.01$} & 
\multicolumn{1}{c|}{$-0.89\pm0.11$} & 
\multicolumn{1}{c|}{$-0.43\pm0.36$}  & 
$\sim 1.1\sigma$ \\
\hline

\multicolumn{1}{|c|}{DESY5(low-$z$+DES-SN)} & 
\multicolumn{1}{c|}{$66.63\pm0.73$} & 
\multicolumn{1}{c|}{$-19.392\pm0.018$} &
\multicolumn{1}{c|}{$0.322^{+0.0064}_{-0.0072}$} & 
\multicolumn{1}{c|}{$-0.79\pm0.067$} & 
\multicolumn{1}{c|}{$-0.72\pm0.28$}  & 
$\sim 2.9\sigma$ \\
\hline

\multicolumn{7}{|c|}{Planck CMB + PantheonPlus + } \\
\hline

\multicolumn{1}{|c|}{DESI BAO DR1} & 
\multicolumn{1}{c|}{$68.0^{+0.84}_{-0.71}$} & 
\multicolumn{1}{c|}{$-19.406^{+0.024}_{-0.019}$} &
\multicolumn{1}{c|}{$0.308\pm0.0076$} & 
\multicolumn{1}{c|}{$-0.83^{+0.059}_{-0.070}$} & 
\multicolumn{1}{c|}{$-0.70^{+0.30}_{-0.27}$}  & 
$\sim 2.5\sigma$ \\
\hline

\multicolumn{1}{|c|}{eBOSS BAO DR16} & 
\multicolumn{1}{c|}{$67.3^{+0.61}_{-0.71}$} & 
\multicolumn{1}{c|}{$-19.429^{+0.017}_{-0.020}$} &
\multicolumn{1}{c|}{$0.316\pm0.0075$} & 
\multicolumn{1}{c|}{$-0.88\pm0.069$} & 
\multicolumn{1}{c|}{$-0.45^{+0.29}_{-0.22}$}  & 
$\sim 1.7\sigma$ \\
\hline

\end{tabular}
\end{flushleft}
\end{table*}

\begin{table*}[!htbp]
\begin{flushleft}
\caption{Constraints on $w_0w_a$CDM model with and without DESI Y1/Y3 BAO before and after rebuilding $a_B$ consistency.}
\label{tab:w0waCDMcorrection}
\renewcommand{\arraystretch}{1.2}
\begin{tabular}{|ccccccc|}
\hline

\multicolumn{7}{|c|}{Planck CMB +}\\
\hline

\multicolumn{1}{|c|}{Datasets} & \multicolumn{1}{c|}{$H_0$} & \multicolumn{1}{c|}{$M_B$} & \multicolumn{1}{c|}{$\Omega_m$} & \multicolumn{1}{c|}{$w_0$} &  \multicolumn{1}{c|}{$w_a$} & Tensions\\ 
\hline

\multicolumn{1}{|c|}{DESY5 (low-$z$+DES-SN)} & 
\multicolumn{1}{c|}{$67.4^{+1.1}_{-1.0}$} & 
\multicolumn{1}{c|}{$-19.367^{+0.038}_{-0.031}$} &
\multicolumn{1}{c|}{$0.315^{+0.010}_{-0.012}$} & \multicolumn{1}{c|}{$-0.73\pm0.10$} & 
\multicolumn{1}{c|}{$-1.05\pm0.51$}  & 
$\sim 2.4\sigma$\\ 
\hline

\multicolumn{1}{|c|}{low-$z$ fluctuation debiased} & 
\multicolumn{1}{c|}{$79.8\pm7.7$} & 
\multicolumn{1}{c|}{$-18.97^{+0.24}_{-0.19}$} & 
\multicolumn{1}{c|}{$0.224^{+0.033}_{-0.056}$} & \multicolumn{1}{c|}{$-0.64^{+0.24}_{-0.19}$ }& \multicolumn{1}{c|}{$-3.7^{+1.7}_{-2.0}$}  & $\sim 1.5\sigma$\\ 
\hline

\multicolumn{7}{|c|}{Planck CMB + DESI Y1 BAO +}\\  
\hline

\multicolumn{1}{|c|}{Datasets} & \multicolumn{1}{c|}{$H_0$} & \multicolumn{1}{c|}{$M_B$} & \multicolumn{1}{c|}{$\Omega_m$} & \multicolumn{1}{c|}{$w_0$} &  \multicolumn{1}{c|}{$w_a$} & Tensions\\ 
\hline

\multicolumn{1}{|c|}{DESY5} & 
\multicolumn{1}{c|}{$67.25\pm0.72$} & 
\multicolumn{1}{c|}{$-19.370\pm0.017$} &
\multicolumn{1}{c|}{$0.316\pm0.0077$} & \multicolumn{1}{c|}{$-0.73\pm0.069$} & 
\multicolumn{1}{c|}{$-1.01\pm0.32$}  & 
$\sim 3.5\sigma$\\ 
\hline

\multicolumn{1}{|c|}{low-$z$ fluctuations debiased} & 
\multicolumn{1}{c|}{$67.07^{+0.74}_{-0.67}$} & 
\multicolumn{1}{c|}{$-19.372\pm0.017$} &
\multicolumn{1}{c|}{$0.317^{+0.0069}_{-0.0076}$} & 
\multicolumn{1}{c|}{$-0.73^{+0.064}_{-0.074}$} & 
\multicolumn{1}{c|}{$-0.99^{+0.33}_{-0.26}$}  & 
$\sim 3.7\sigma$ \\
\hline

\multicolumn{1}{|c|}{low-$z$ discrepancy debiased} & 
\multicolumn{1}{c|}{$68.75\pm0.76$} & 
\multicolumn{1}{c|}{$-19.370\pm0.018$} &
\multicolumn{1}{c|}{$0.301^{+0.0062}_{-0.0073}$} & 
\multicolumn{1}{c|}{$-0.91\pm0.067$} & 
\multicolumn{1}{c|}{$-0.49\pm0.28$}  & 
$\sim 1.6\sigma$ \\
\hline

\multicolumn{1}{|c|}{low-$z$ fully debiased} & 
\multicolumn{1}{c|}{$68.68\pm0.80$} & 
\multicolumn{1}{c|}{$-19.371\pm0.019$} &
\multicolumn{1}{c|}{$0.302\pm0.0079$} & 
\multicolumn{1}{c|}{$-0.91\pm0.070$} & 
\multicolumn{1}{c|}{$-0.49^{+0.30}_{-0.26}$}  & 
$\sim 1.5\sigma$\\
\hline

\multicolumn{7}{|c|}{DESI Y1 BAO + }\\ 
\hline

\multicolumn{1}{|c|}{Datasets} &  \multicolumn{1}{c|}{$r_d$}  & \multicolumn{1}{c|}{$M_B$}  & \multicolumn{1}{c|}{$\Omega_m$} & \multicolumn{1}{c|}{$w_0$} & \multicolumn{1}{c|}{$w_a$} & Tensions\\ 
\hline

\multicolumn{1}{|c|}{DESY5} & 
\multicolumn{1}{c|}{$146.6\pm1.5$} & 
\multicolumn{1}{c|}{$-19.365\pm0.014$} & 
\multicolumn{1}{c|}{$0.329^{+0.02}_{-0.015}$} &
\multicolumn{1}{c|}{$-0.734^{+0.087}_{-0.10}$} &
\multicolumn{1}{c|}{$-1.22\pm0.68$}& 
$\sim 2.4\sigma$ \\ 
\hline

\multicolumn{1}{|c|}{low-$z$ fluctuations debiased} & 
\multicolumn{1}{c|}{$146.6\pm1.5$ }& 
\multicolumn{1}{c|}{$-19.365\pm0.014$}  &
\multicolumn{1}{c|}{$0.328^{+0.019}_{-0.015}$} & 
\multicolumn{1}{c|}{$-0.733^{+0.087}_{-0.10}$} & 
\multicolumn{1}{c|}{$-1.22\pm0.67$} &  
$\sim 2.4\sigma$ \\ 
\hline

\multicolumn{1}{|c|}{low-$z$ discrepancy debiased} & 
\multicolumn{1}{c|}{$150.1\pm1.6$} &
\multicolumn{1}{c|}{$-19.417\pm0.014$}  &
\multicolumn{1}{c|}{$0.295^{+0.032}_{-0.017}$} & 
\multicolumn{1}{c|}{$-0.957^{+0.068}_{-0.08}$} &
\multicolumn{1}{c|}{$-0.11^{+0.74}_{-0.50}$} &  
$\sim 0.5\sigma$ \\ 
\hline

\multicolumn{1}{|c|}{low-$z$ fully debiased} &
\multicolumn{1}{c|}{$150.1\pm1.5$ }& 
\multicolumn{1}{c|}{$-19.418\pm0.014$}  & 
\multicolumn{1}{c|}{$0.282^{+0.05}_{-0.01}$} & 
\multicolumn{1}{c|}{$-0.953^{+0.066}_{-0.078}$} & 
\multicolumn{1}{c|}{$0.0^{+0.97}_{-0.58}$} & 
$\sim 0.5\sigma$ \\ 
\hline

\multicolumn{7}{|c|}{Planck CMB + DESI Y3 BAO +}                      \\  \hline

\multicolumn{1}{|c|}{Datasets} & \multicolumn{1}{c|}{$H_0$} & \multicolumn{1}{c|}{$M_B$} & \multicolumn{1}{c|}{$\Omega_m$} & \multicolumn{1}{c|}{$w_0$} &  \multicolumn{1}{c|}{$w_a$} & Tensions\\ 
\hline

\multicolumn{1}{|c|}{DESY5} & 
\multicolumn{1}{c|}{$66.97\pm0.66$} & 
\multicolumn{1}{c|}{$-19.379\pm0.017$} &
\multicolumn{1}{c|}{$0.318\pm0.0069$} & 
\multicolumn{1}{c|}{$-0.76\pm0.06$} & 
\multicolumn{1}{c|}{$-0.86\pm0.27$}  & 
$\sim 3.7\sigma$ \\
\hline

\multicolumn{1}{|c|}{low-$z$ fluctuations debiased} & 
\multicolumn{1}{c|}{$66.76^{+0.74}_{-0.66}$} & 
\multicolumn{1}{c|}{$-19.382\pm0.017$} &
\multicolumn{1}{c|}{$0.320^{+0.0066}_{-0.0075}$} & 
\multicolumn{1}{c|}{$-0.74\pm0.068$} & 
\multicolumn{1}{c|}{$-0.88^{+0.32}_{-0.24}$}  & 
$\sim 3.5\sigma$ \\
\hline

\multicolumn{1}{|c|}{low-$z$ discrepancy debiased} & 
\multicolumn{1}{c|}{$68.30\pm0.69$} & 
\multicolumn{1}{c|}{$-19.379\pm0.017$} &
\multicolumn{1}{c|}{$0.305\pm0.0067$} & 
\multicolumn{1}{c|}{$-0.90^{+0.057}_{-0.065}$} & 
\multicolumn{1}{c|}{$-0.45^{+0.30}_{-0.23}$}  & 
$\sim 1.6\sigma$ \\
\hline

\multicolumn{1}{|c|}{low-$z$ fully debiased} & 
\multicolumn{1}{c|}{$68.39\pm0.78$} & 
\multicolumn{1}{c|}{$-19.377\pm0.017$} &
\multicolumn{1}{c|}{$0.305^{+0.0061}_{-0.0073}$} & 
\multicolumn{1}{c|}{$-0.91\pm0.063$} & 
\multicolumn{1}{c|}{$-0.43^{+0.28}_{-0.23}$}  & 
$\sim 1.6\sigma$\\
\hline

\multicolumn{7}{|c|}{DESI Y3 BAO + }\\ 
\hline
\multicolumn{1}{|c|}{Datasets} &  \multicolumn{1}{c|}{$r_d$}  & \multicolumn{1}{c|}{$M_B$}  & \multicolumn{1}{c|}{$\Omega_m$} & \multicolumn{1}{c|}{$w_0$} & \multicolumn{1}{c|}{$w_a$} & Tensions\\ 
\hline

\multicolumn{1}{|c|}{DESY5} & 
\multicolumn{1}{c|}{$146.1\pm1.6$} & 
\multicolumn{1}{c|}{$-19.368\pm0.014$} & 
\multicolumn{1}{c|}{$0.317^{+0.039}_{-0.018}$} &
\multicolumn{1}{c|}{$-0.787^{+0.074}_{-0.088}$} &
\multicolumn{1}{c|}{$-0.69\pm0.69$}& 
$\sim 2.0\sigma$ \\
\hline

\multicolumn{1}{|c|}{low-$z$ fluctuations debiased} & 
\multicolumn{1}{c|}{$146.1\pm1.6$ }& 
\multicolumn{1}{c|}{$-19.369\pm0.014$}  &
\multicolumn{1}{c|}{$0.308^{+0.047}_{-0.013}$} & 
\multicolumn{1}{c|}{$-0.788^{+0.075}_{-0.087}$} & 
\multicolumn{1}{c|}{$-0.60^{+0.65}_{-0.78}$} &  
$\sim 1.9\sigma$ \\
\hline

\multicolumn{1}{|c|}{low-$z$ discrepancy debiased} & 
\multicolumn{1}{c|}{$149.6\pm1.6$} &
\multicolumn{1}{c|}{$-19.418\pm0.013$}  &
\multicolumn{1}{c|}{$0.238^{+0.098}_{-0.024}$} & 
\multicolumn{1}{c|}{$-0.943^{+0.069}_{-0.11}$} &
\multicolumn{1}{c|}{$0.46^{+0.66}_{-0.27}$} &  
$\sim 0.8\sigma$ \\ 
\hline

\multicolumn{1}{|c|}{low-$z$ fully debiased} &
\multicolumn{1}{c|}{$149.7\pm1.6$ }& 
\multicolumn{1}{c|}{$-19.419\pm0.013$}  & 
\multicolumn{1}{c|}{$0.248^{+0.086}_{-0.023}$} & 
\multicolumn{1}{c|}{$-0.954^{+0.069}_{-0.099}$} & 
\multicolumn{1}{c|}{$0.42^{+0.69}_{-0.29}$} & 
$\sim 0.7\sigma$ \\ 
\hline 

\end{tabular}
\end{flushleft}
\end{table*}

\bibliography{ref}

\end{document}